\documentclass[journal]{IEEEtran}
\usepackage{amsmath,amsfonts}
\usepackage{algorithmic}
\usepackage{algorithm}
\usepackage{array}
\usepackage[caption=false,font=normalsize,labelfont=sf,textfont=sf]{subfig}
\usepackage{textcomp}
\usepackage{stfloats}
\usepackage{url}
\usepackage{verbatim}
\usepackage{graphicx}
\usepackage{arydshln}
\usepackage{cite}
\usepackage{float}
\usepackage{booktabs}
\usepackage{svg}
\usepackage{graphicx,array}
\usepackage{authblk}
\usepackage{booktabs}
\usepackage{makecell}
\usepackage[font=footnotesize, labelsep=period]{caption}
\usepackage{float}
\usepackage[colorlinks=true, citecolor=blue, linkcolor=black, urlcolor=cyan]{hyperref}
\begin{document}

\title{A Dual-Bearing Magnetorheological Grease Clutch with Intention-Based Demagnetization for Wearable Haptic Feedback}

\author{
    \IEEEauthorblockN{
        Zhongyuan Kong,
        Lei Li,
        Erwin Ang Tien Yew,
        Zirui Chen,
        Wenbo Li,
        Shiwu Zhang,
        Jian Yang\IEEEauthorrefmark{1},
        Shuaishuai Sun\IEEEauthorrefmark{1}
    }
    
\thanks{
    This work was supported by National Natural Science Foundation of China (Grant No. U21A20119 and 52105081), Major Project of Anhui Province's Science and Technology Innovation Breakthrough Plan (Grant No. 202423h08050003), Anhui's Key R\&D Program of China (Grant No. 202104a05020009), USTC start-up funding (Grant No. KY2090000067), and the Fundamental Research Funds for the Central Universities (Grant No. YD2090002019).

    This work involved human subjects. All experimental procedures and protocols were approved by the Institutional Review Board of the First Affiliated Hospital of the University of Science and Technology of China (Approval No. 2023KY239). Written informed consent was obtained from all participants prior to the experiments.
    
    Zhongyuan Kong, Lei Li, Erwin Ang Tien Yew, Zirui Chen, Wenbo Li, Shiwu Zhang and Shuaishuai Sun are with the CAS Key Laboratory of Mechanical Behavior and Design of Materials and Institute of Humanoid Robots, School of Engineering Sciences, University of Science and Technology of China, Hefei, Anhui, 230026, China.
    
    Jian Yang is with the School of Electrical Engineering and Automation, Anhui University, Hefei, Anhui, 230601, China.
    
    *Corresponding author emails: 20152@ahu.edu.cn; sssun@ustc.edu.cn.
    }
}

\maketitle

\begin{abstract}
This paper presents the design, modeling, and control of a dual-bearing magnetorheological grease (MRG) clutch for wearable haptic feedback. Compared with conventional MR fluid devices, the proposed clutch avoids leakage-related reliability degradation while achieving high torque density in a compact structure. To provide physical insight into the torque-generation mechanism, a physics-inspired interpretive model is introduced to capture the dominant relationship among excitation current, magnetic-field evolution in the bearing gaps, and clutch locking torque. To mitigate the undesirable ``sticky'' sensation caused by passive bidirectional braking, an intention-based control strategy with active demagnetization is further developed to enable smoother release during human--robot interaction. Experimental characterization shows that the proposed clutch achieves a maximum locking torque of 43.42\,N$\cdot$m at 1.3\,A and a torque-to-mass ratio of 96.5\,N$\cdot$m/kg. Bench tests, replay validation, teleoperation experiments, and user studies indicate that the proposed approach accelerates clutch release and improves perceived release transparency and contact-to-release smoothness, while maintaining effective multi-level kinesthetic rendering.
\end{abstract}


\section{Introduction}

\IEEEPARstart{T}{eleoperated} robotic systems have attracted increasing attention for remote operations in hazardous and unstructured environments, especially in space exploration~\cite{Guo2024}. In such scenarios, haptic feedback is essential for improving interaction safety, manipulation precision, and operator situational awareness, and prior studies have shown that both the presence and the quality of haptic feedback can materially affect teleoperation performance and operator interaction behavior~\cite{Huang2025,Feizi2023,MohandOusaid2020,Cabibihan2021,Nitsch2013,Wildenbeest2013}. Wearable exoskeleton-based interfaces are therefore regarded as a promising solution, since they can provide intuitive motion mapping while directly rendering interaction forces to the human operator~\cite{Ma2015,Buongiorno2018,Varghese2020}.

Recent wearable-haptics literature has highlighted persistent tradeoffs among force capability, backdrivability, portability, and wearability in hand- and upper-limb-oriented interfaces~\cite{Wang2019,Pacchierotti2017,duPasquier2024}. Although these approaches can generate controllable force output, they often face practical limitations in wearable applications. Cable-driven systems usually require remote transmission layouts and are difficult to integrate with compact joint-level feedback. Pneumatic systems improve compliance and wearability, yet their response delay and air supply dependence reduce force transparency. Motor-reducer-based devices can provide large torque, but their limited backdrivability, structural complexity, and power consumption often compromise transparency and portability~\cite{Forouhar2024,Su2021,Nassour2021,Li2019,Porcini2020}. These issues motivate the exploration of semi-active haptic mechanisms that are intrinsically safe, compact, and energy efficient. In teleoperation and wearable haptics, passive or variable-impedance feedback strategies have been explored because of their safety and stability advantages, although they often face limitations in feedback richness, directionality, or release behavior~\cite{Pacchierotti2015,Zhang2019,Pacchierotti2024}.

Among semi-active technologies, electrorheological materials, electroadhesion clutches, and MR devices are representative solutions. However, electrorheological devices typically require very high electric fields while offering limited yield stress, which constrains torque density in compact rotary joints. Electroadhesion clutches feature low holding-power consumption, but their output is sensitive to surface condition and environmental factors, which undermines repeatability in wearable systems. MR devices are attractive because of their fast response, simple excitation, and passive energy-dissipative nature~\cite{Carlson2000,DeVicente2011}. Nevertheless, conventional MR-fluid-based clutches still face two major limitations. First, leakage and sedimentation during long-term use degrade output stability and reliability. Second, MR clutches inherently generate bidirectional braking torque. In contact tasks, the user is expected to feel resistance when moving toward a virtual object, but to disengage smoothly when moving away. A conventional MR clutch cannot distinguish these two intentions and may continue resisting the release motion, resulting in an undesirable ``sticky'' sensation. Therefore, the core challenge addressed in this work is not only to provide compact and high-torque semi-active haptic rendering, but also to mitigate the delayed and sticky release sensation during disengagement.

To address the material limitation of MR fluids, MRG has emerged as a promising alternative~\cite{Park2010}. Compared with MR fluids, MRG preserves field-dependent rheological tunability while effectively suppressing leakage during long-term operation~\cite{Raj2021,Park2010}. Motivated by this advantage, this paper develops a wearable haptic clutch based on MRG. The proposed device adopts a symmetric dual-bearing configuration, in which two MR bearings are arranged on both sides of the excitation coil to increase magnetic-field utilization and torque capacity. This structure enables a compact and leakage-free clutch with high torque density, making it suitable for high-load upper-limb joints such as the shoulder and elbow.

Beyond device realization, this work also emphasizes model-based understanding of torque generation. A physics-inspired interpretive model is introduced to capture the dominant relationship among the excitation current, the magnetic-field evolution in the bearing gaps, the field-dependent rheological transition of the MRG, and the resulting clutch torque. Rather than serving as a fully predictive material-level constitutive model, this interpretive model is introduced to explain the dominant torque-generation mechanism of the dual-bearing clutch, account for the experimentally observed monotonic increase and saturation trend of locking torque, and clarify the torque enhancement introduced by the symmetric structure. In this way, the proposed clutch is not only experimentally characterized, but also physically interpreted beyond purely empirical current--torque fitting. For control implementation, however, a compact experimental current--torque mapping is still adopted for real-time command generation due to its simplicity and calibration accuracy.

To improve directional transparency during contact transitions, an intention-based control strategy is further developed. Thin-film force sensors are used to infer whether the operator intends to approach or disengage from the virtual object. Once a release intention is detected, an active demagnetization process is triggered to rapidly suppress the residual magnetic effect and release the clutch. By combining high-torque MRG hardware with intention-aware control, the proposed system improves release transparency and transition smoothness while preserving the safety and efficiency of passive haptic rendering.

The main contributions of this work are summarized as follows:

\begin{itemize}

\item \textbf{High-Performance Device Design:}
A compact and leakage-free MRG clutch is developed using an optimized magnetic circuit and a symmetric dual-bearing architecture. The prototype achieves a maximum locking torque of 43.42~N$\cdot$m at 1.3~A, together with a torque-to-mass ratio (TMR) of 96.5~N$\cdot$m/kg, a torque-to-volume ratio (TVR) of 3.98$\times$10$^{5}$~N$\cdot$m/m$^{3}$, and a torque-to-power ratio (TPR) of 4.28~N$\cdot$m/W.

\item \textbf{Physics-Inspired Interpretive Modeling:}
A physics-inspired interpretive model of the dual-bearing MRG clutch is established to capture the dominant relationship among excitation current, magnetic-field evolution in the bearing gaps, and locking torque generation. The model provides physical insight into the torque amplification effect of the symmetric dual-bearing structure and the observed saturation behavior of the clutch.

\item \textbf{Intention-Based Release Enhancement:}
An intention-based active demagnetization strategy is proposed to mitigate the sticky effect caused by passive bidirectional braking, thereby improving release promptness and contact-to-release smoothness in wearable haptic interaction.

\end{itemize}

The remainder of this paper is organized as follows. Section~II presents the mechanical design, physics-inspired interpretive model, magnetic-field simulation, and torque characterization of the proposed clutch. Section~III introduces the hardware driver and the intention-based control implementation. Section~IV reports the experimental validation, including bench tests, human-subject studies, and teleoperation tasks.

\section{Design and Modeling of the MRG clutch}

\subsection{Mechanical and Device Design}

The internal structure of the proposed MRG clutch is illustrated in Fig.~1(a). The stator consists of a sealed cup, a stator ring, and aluminum-alloy base~1, which are fixed to the parent link of the exoskeleton joint. The rotor consists of a rotor ring and aluminum-alloy base~2, which are fixed to the child link. The excitation coil and two MR bearings are arranged between the rotor and stator. To enhance magnetic-field utilization and torque capacity, a symmetric dual-bearing configuration in the form of ``MR bearing--coil--MR bearing'' is adopted.

The key component of the clutch is the MR bearing, which consists of a deep-groove ball bearing filled with MRG in its internal gap spaces. The MRG is prepared by mixing high-viscosity lubricating grease with iron powder of approximately 0.5\,$\mu$m particle size. During preparation, the grease is first heated in a water bath to 80\,\textdegree C and stirred at 500\,rpm for 10\,min. The iron powder is then added, and the mixture is stirred at 800\,rpm for 60\,min at 80\,\textdegree C until the particles are uniformly dispersed. To fabricate the MR bearing, the prepared MRG is injected into an opened deep-groove ball bearing. The bearing is then placed in a vacuum oven for 5\,min to remove trapped air and facilitate full penetration of the grease into the bearing gaps. After several repeated filling operations, the bearing cover is sealed.

The operating principle of the MR bearing is shown in Fig.~1(b). Without an applied magnetic field, the magnetic particles remain randomly distributed in the grease, resulting in low off-state resistance. Under this condition, the MR bearings rotate freely, and the wearer experiences little haptic resistance. Once a magnetic field is applied, the magnetic particles rapidly form chain-like structures along the field direction, thereby increasing the shear resistance of the MRG and generating a controllable locking torque. By adjusting the excitation current, the magnetic flux density in the bearing gaps can be regulated, and the locking torque can therefore be continuously tuned over the practical operating range. As the magnetic field gradually approaches saturation, further increases in current lead to only marginal torque growth. In this state, the rotational motion of the MR bearing is strongly suppressed, enabling the clutch to provide substantial resistance against unintended or excessive joint motion and thereby improving the safety of the teleoperation system.

\begin{figure*}[tbp]
    \centering
    \includegraphics[width=\linewidth]{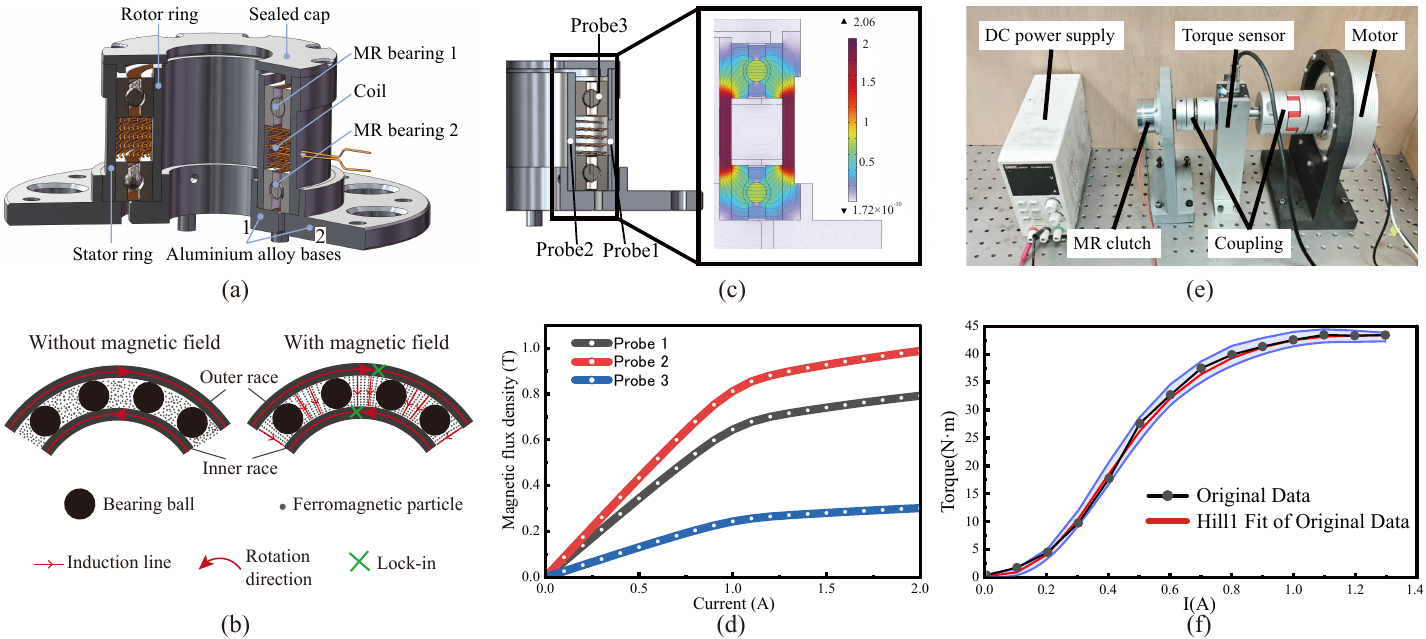}
    \caption{Structure, working principle, and magnetic field analysis of the MRG clutch. (a) Cutaway view of MRG clutch structure. (b) Principle of MR bearings. (c) Magnetic flux density simulation with 1.3 A current input. The illustration displays the symmetric magnetic circuit formed in the construction of the MRG clutch and the sites of the three probes. (d) Magnetic flux density of the three probes under various currents. (e) MRG clutch testing apparatus and platform. (f) Measured locking torque of the proposed clutch under different excitation currents and the corresponding Hill1-function fitting curve, the error band indicates the maximum observed deviation across repeated trials.}
\end{figure*}

\subsection{Physics-Inspired Interpretive Model of the Dual-Bearing MRG Clutch}

To provide a physically interpretable description of the torque-generation mechanism of the proposed dual-bearing MRG clutch, a physics-inspired interpretive model is established. The model is not intended to serve as a fully predictive material-level constitutive model or a real-time control law. Instead, it is used to capture the dominant relationships among excitation current, magnetic-field evolution in the bearing gaps, field-dependent rheological transition of the MRG, and clutch torque generation.

As shown in Fig.~1(a), the proposed clutch adopts a symmetric configuration in the form of ``MR bearing--coil--MR bearing,'' where the two MR bearings are activated by the same magnetic circuit. During the locking-torque characterization considered in this work, the clutch operates at relatively low speed. Therefore, the dominant output torque can be interpreted as arising mainly from the field-dependent resistive behavior of the MRG in the bearing gaps.

To describe this dominant behavior in a compact way, a Bingham-type relation is adopted as an effective representation of the activated MRG region:
\begin{equation}
\tau = \tau_y(B_{\mathrm{gap}}) + \eta_{\mathrm{eq}} \dot{\gamma},
\label{eq:bingham_interpretive}
\end{equation}
where $\tau$ is the equivalent shear stress, $\tau_y(B_{\mathrm{gap}})$ is the effective field-dependent yield stress, $\eta_{\mathrm{eq}}$ is the equivalent viscosity, and $\dot{\gamma}$ is the equivalent shear rate. Here, (\ref{eq:bingham_interpretive}) is introduced as a physics-inspired representation of the dominant field-dependent resistive mechanism, rather than a full constitutive identification of the MRG material.

Because the local gap geometry and magnetic-field distribution inside the MR bearing are spatially non-uniform, the distributed interaction is described using lumped effective parameters. The equivalent shear rate is approximated as
\begin{equation}
\dot{\gamma} \approx \frac{R_{\mathrm{eff}}\omega_{\mathrm{rel}}}{h_{\mathrm{eq}}},
\label{eq:shearrate_interpretive}
\end{equation}
where $\omega_{\mathrm{rel}}$ is the relative angular velocity between the rotor and stator, $R_{\mathrm{eff}}$ is the effective moment arm of the activated MRG region, and $h_{\mathrm{eq}}$ is the effective shear-gap thickness.

Based on this approximation, the torque generated by a single MR bearing can be expressed as
\begin{equation}
T_b=
\Gamma_b N_b A_{\mathrm{eff}} R_{\mathrm{eff}}
\left[
\tau_y(B_{\mathrm{gap}})
+\eta_{\mathrm{eq}}\frac{R_{\mathrm{eff}}}{h_{\mathrm{eq}}}\omega_{\mathrm{rel}}
\right]
+T_{\mathrm{off},b},
\label{eq:Tb_interpretive}
\end{equation}
where $N_b$ is the number of effective magnetically loaded rolling contacts, $A_{\mathrm{eff}}$ is the effective activated shear area associated with each contact, $\Gamma_b \in (0,1]$ is an effectiveness factor accounting for non-uniform magnetic activation, and $T_{\mathrm{off},b}$ denotes the off-state drag torque caused by grease viscosity, seal friction, and other mechanical parasitics.

Since the clutch employs two symmetrically arranged MR bearings driven by the same excitation coil, the total clutch torque can be written as
\begin{equation}
T_c = 2T_b.
\label{eq:Tc_interpretive}
\end{equation}
Defining a lumped effective coefficient
\begin{equation}
K_c = 2\Gamma_b N_b A_{\mathrm{eff}} R_{\mathrm{eff}},
\label{eq:Kc_interpretive}
\end{equation}
the total clutch torque is rewritten in a compact form as
\begin{equation}
T_c=
K_c
\left[
\tau_y(B_{\mathrm{gap}})
+\eta_{\mathrm{eq}}\frac{R_{\mathrm{eff}}}{h_{\mathrm{eq}}}\omega_{\mathrm{rel}}
\right]
+T_{\mathrm{off}},
\label{eq:Tc_full_interpretive}
\end{equation}
where $T_{\mathrm{off}}=2T_{\mathrm{off},b}$ is the off-state torque of the dual-bearing clutch.

The magnetic-field simulation in Fig.~1(c)--(d) shows that the magnetic flux density in the bearing-gap region increases monotonically with excitation current and gradually approaches saturation at higher current levels. Motivated by this trend, the current-to-field relationship is represented by a monotonic saturating mapping
\begin{equation}
B_{\mathrm{gap}} = g(I),
\label{eq:BI_general_interpretive}
\end{equation}
where $I$ is the excitation current and $g(\cdot)$ is an effective mapping used to describe the dominant evolution of the gap field. A compact expression is adopted as
\begin{equation}
B_{\mathrm{gap}}(I)=
B_0+\left(B_{\mathrm{sat}}-B_0\right)\left(1-e^{-k_I I}\right),
\label{eq:BI_sat_interpretive}
\end{equation}
where $B_0$ is the residual gap flux density, $B_{\mathrm{sat}}$ is the effective saturation flux density, and $k_I$ is an effective current-to-field coefficient. These parameters are interpreted here as descriptive quantities of the current--field evolution rather than exact electromagnetic identification constants.

Similarly, the field-dependent yield stress is represented by an effective saturating relation
\begin{equation}
\tau_y(B_{\mathrm{gap}})
=
\tau_{\max}\left(1-e^{-k_B B_{\mathrm{gap}}}\right),
\label{eq:tauB_interpretive}
\end{equation}
where $\tau_{\max}$ denotes the effective saturated yield stress and $k_B$ is an effective field-sensitivity coefficient. This expression is introduced to reflect the observed increase and eventual saturation of the clutch torque with increasing excitation current.

Combining (\ref{eq:Tc_full_interpretive})--(\ref{eq:tauB_interpretive}), the current- and speed-dependent clutch torque can be interpreted as
\begin{equation}
\begin{aligned}
T_c(I,\omega_{\mathrm{rel}})
={}&
K_c
\Big[
\tau_{\max}\big(1-e^{-k_B g(I)}\big)
+\eta_{\mathrm{eq}}\frac{R_{\mathrm{eff}}}{h_{\mathrm{eq}}}\omega_{\mathrm{rel}}
\Big] \\
&+T_{\mathrm{off}}.
\end{aligned}
\label{eq:T_Iw_interpretive}
\end{equation}

For the low-speed locking condition considered in this work, the viscous contribution is much smaller than the field-induced yield contribution. Therefore, the above expression can be simplified into a compact saturating form:
\begin{equation}
T_{\mathrm{lock}}(I)
\approx
A_c\big(1-e^{-k_B g(I)}\big)+T_{\mathrm{off}},
\label{eq:T_lock_interpretive}
\end{equation}
where
\begin{equation}
A_c = K_c \tau_{\max}
\label{eq:Ac_interpretive}
\end{equation}
is an effective torque coefficient lumping the dominant geometric and rheological contributions.

Since the experiments in this work focus on low-speed locking characterization, (\ref{eq:T_lock_interpretive}) is the main interpretive expression used in the following discussion. It provides a concise explanation of the main experimental observations of the proposed clutch. First, the locking torque increases monotonically with excitation current because the magnetic field enhances the effective yield stress of the MRG. Second, the current--torque relationship gradually saturates at higher current levels because both the gap-field evolution and the field-dependent yield stress are saturating processes. Third, the dual-bearing configuration increases the overall output torque by effectively stacking two activated MR interfaces within the same clutch structure.

It should be emphasized that the proposed model is physics-inspired and interpretive in nature. Its purpose is to explain the dominant torque-generation mechanism and provide design-level physical insight into the effects of magnetic activation, structural symmetry, and saturation behavior. For real-time control implementation, however, a compact experimental current--torque mapping is still preferred due to its simplicity and calibration accuracy. Therefore, for practical current command generation, the measured current--torque data are further fitted by a Hill1 function in Section~II-D.

\subsection{Magnetic Field Simulation and Model Support}

Magnetic-field simulation was conducted to visualize the field distribution and verify the feasibility of the proposed magnetic circuit. In COMSOL Multiphysics, the material properties of all components were assigned according to their actual physical characteristics, and the excitation current of the coil was swept from 0 to 2.0\,A. As shown in Fig.~1(c), three probes were defined in the model to monitor the magnetic flux density at representative locations. Probes~1 and~2 were placed in the stator ring and rotor ring, respectively, while Probe~3 was placed in the MRG-bearing gap region, where the field directly affects the rheological state of the grease. The simulation is used to support the plausibility of the monotonic saturating current-to-field trend, rather than to quantitatively predict the measured torque curve.

As shown in Fig.~1(d), the magnetic flux density measured by all three probes increases with the excitation current. In particular, the response of Probe~3 confirms that the magnetic field acting on the MR bearings can be effectively regulated by the coil current. Meanwhile, the slope gradually decreases as the current rises, indicating the onset of magnetic saturation. The simulation results show that the magnetic circuit approaches saturation at approximately 1.3\,A, which is consistent with the experimentally observed saturation behavior of the locking torque.

Therefore, the COMSOL results not only verify the effectiveness of the magnetic design, but also support the use of a monotonic saturating mapping $g(I)$ in the proposed interpretive model. In other words, the simulation is used here primarily to support the physical plausibility of the current-to-field evolution assumed in Section~II-B, rather than to establish a fully predictive electromagnetic identification model.

\subsection{Experimental Torque Characterization and Control-Oriented Fitting}

To further investigate the relationship between excitation current and clutch torque, an experimental setup was established, as shown in Fig.~1(e). The test bench consists of three main components. A servo motor with a rated torque of 100\,N$\cdot$m is installed on the right side to drive the system. A torque sensor with a range of 100\,N$\cdot$m and an accuracy of 0.1\% is mounted at the center to measure the transmitted torque. The MRG clutch is fixed on the left side. These components are mounted on an optical platform using aluminum-alloy brackets and connected through couplings to ensure alignment and reliable torque transmission. During the experiment, the motor speed is regulated by a controller, the torque signal is acquired through the sensor, and the clutch current is supplied by a programmable DC power source.

The measured relationship between excitation current and locking torque is shown in Fig.~1(f). Each data point represents the average value obtained from repeated trials. When the excitation current is 0\,A, the clutch exhibits an off-state torque of approximately 0.2\,N$\cdot$m. As the current increases, the locking torque rises monotonically and reaches 43.42\,N$\cdot$m at 1.3\,A. Beyond this point, additional increases in current produce only negligible torque growth, indicating that the clutch has entered a near-saturation regime. Therefore, the effective operating range considered in this study is 0--1.3\,A.

The measured torque-current relationship is consistent with the dominant trend described by the proposed physics-inspired interpretive model, namely, monotonic torque growth at low current and gradual saturation at higher current levels. However, the model introduced in Section~II-B is intended primarily for mechanism interpretation and design understanding rather than precise quantitative fitting.

For real-time control implementation, a compact algebraic mapping is preferred over the interpretive model because of its simplicity and calibration accuracy. Therefore, the experimental current--torque data are further fitted using a Hill1 function:
\begin{equation}
T_{\mathrm{fit}}(I)=
44.026\frac{I^{2.671}}{0.465^{2.671}+I^{2.671}},
\label{eq:Hill1}
\end{equation}
where $I$ denotes the excitation current and $T_{\mathrm{fit}}$ denotes the fitted locking torque. This control-oriented mapping is used for practical current command generation in the proposed haptic rendering strategy.

To evaluate the fidelity of the fitting model, repeated experiments were conducted to quantify the deviation between the measured torque and the fitted curve. The error band illustrated in Fig. 1(f) denotes the maximum observed deviation across all trials, representing a conservative estimate of the model uncertainty. Quantitatively, based on the Hill1 fitting curve, the calculated mean absolute error (MAE), root mean square error (RMSE), and normalized RMSE (nRMSE) are 1.807 N$\cdot$m, 2.039 N$\cdot$m, and 4.5\%, respectively. The close alignment between the experimental data and the fitted model demonstrates sufficient accuracy to support subsequent haptic control law development.

\section{System Integration and Control}

\subsection{Teleoperation System Architecture}

\begin{figure*}[tbp]
    \centering
    \includegraphics[width=\linewidth]{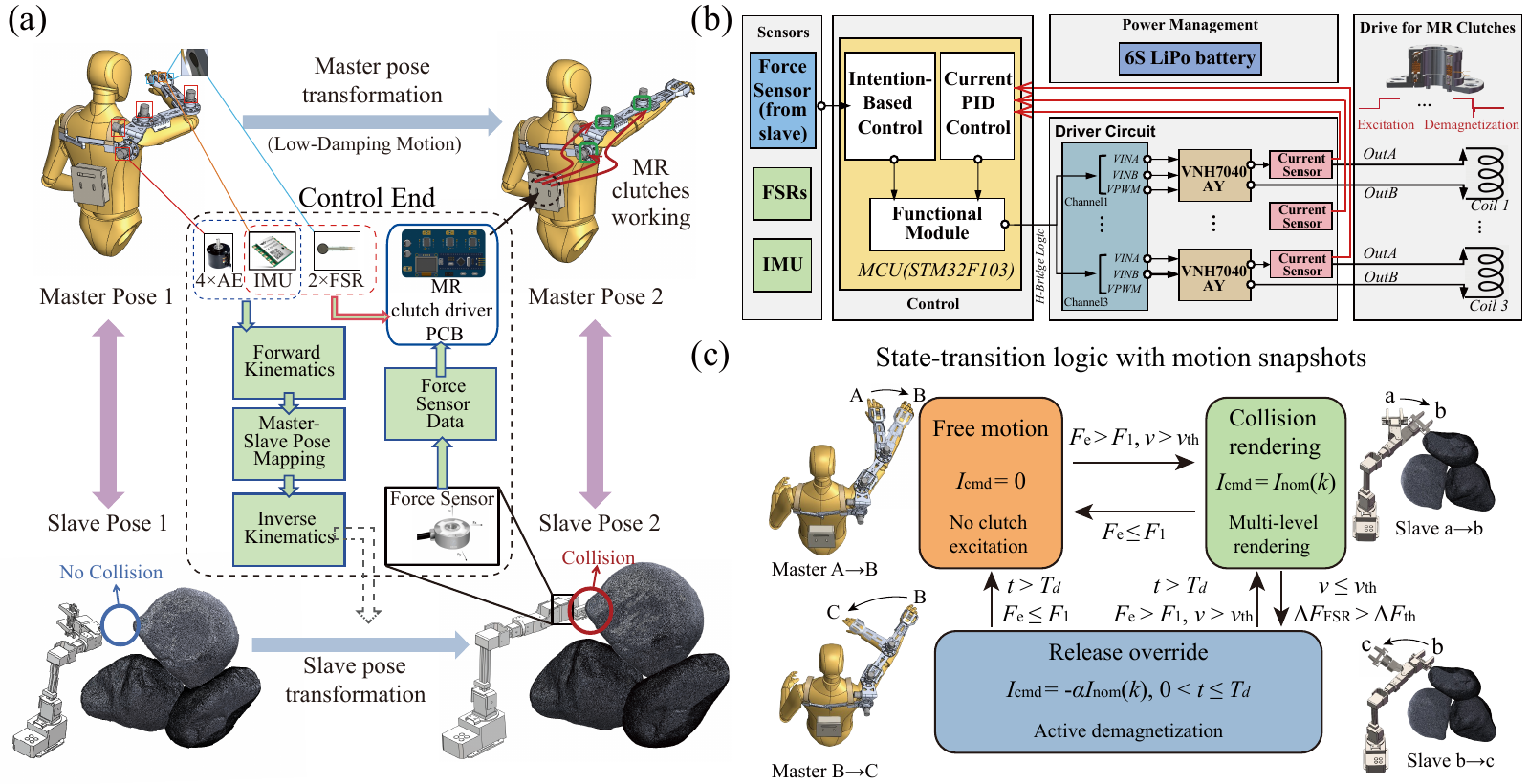}
    \caption{Architecture of the integrated driving system and intention-based control strategy. (a) Teleoperation and haptic-feedback workflow. (b) Hardware architecture of the MRG clutch driver, illustrating the signal flow from multi-sensor acquisition to the H-bridge circuit designed for bidirectional current control. (c) Control logic schematic showing the state transitions between collision rendering (excitation) and intention-based release (demagnetization).}
    \label{fig:system_architecture}
\end{figure*}

The teleoperation system comprises three primary components: the master end, the slave end, and the control end, as shown in Fig.~\ref{fig:system_architecture}(a).

The master end is a single-arm upper-limb exoskeleton equipped with three actuated joints using MRG clutches and one passive shoulder internal/external rotation joint. The exoskeleton incorporates four DOFs corresponding to shoulder abduction/adduction, flexion/extension, internal/external rotation, and elbow flexion/extension~\cite{yan2014}. To avoid structural interference between the exoskeleton and the operator's body, the shoulder internal/external rotation joint is designed as a passive DOF equipped only with an angular encoder (AE). The other three joints integrate both MRG clutches and angular encoders. Consequently, the MRG clutches provide joint-level resistive feedback in the actuated DOFs, while the angular encoders capture joint motion data for all four DOFs. An inertial measurement unit (IMU) is mounted on the dorsum of the operator's hand to acquire hand orientation and motion-state information.

Two thin-film force sensors (FSRs) are integrated into the handle contact surfaces. Together with the IMU, these sensors form a local sensing loop directly connected to the MRG clutch driver PCB. This architecture allows the driver to locally evaluate operator intention and trigger rapid release actions without relying on the main teleoperation loop. The MRG clutch driver module is mounted on the operator's back to improve wearability and motion flexibility.

The slave end employs a 7-DOF robotic arm (K1, Unitree Robotics). A three-dimensional force sensor is mounted at the end-effector to measure remote interaction forces.

At the control end, forward kinematics is applied to process the joint-angle and IMU data into the position and orientation of the master hand relative to the shoulder joint~\cite{Cheng2024}. The kinematic modeling of the upper-limb exoskeleton and the D-H parameter table are provided in Supplementary Fig.~S1 and Supplementary Fig.~S2. The master-hand pose is then mapped to the end-effector pose of the K1 arm, and inverse kinematics is used to drive the slave manipulator.

Meanwhile, the remote interaction force measured at the slave end is converted into a \emph{nominal haptic rendering command}. In the current implementation, this command is realized as a multi-level nominal current input within the validated operating range of the clutch according to the experimentally fitted current--torque relationship in Section~II-D. During normal collision rendering, this nominal current is delivered to the clutch driver to generate the desired resistive feedback. In parallel, a local reflex loop continuously monitors the handle-side FSR signals and IMU-derived motion state. Once a release intention is detected, the local loop temporarily overrides the nominal rendering command and injects an active demagnetization pulse, thereby enabling rapid and transparent disengagement. This local override architecture reduces the dependence of release responsiveness on the communication and control latency of the main teleoperation loop.

\subsection{Control-Oriented Current Command Generation}

The fundamental principle of the MRG clutch relies on the field-induced rheological transformation of the MRG. Under magnetic excitation, the particles align into chain-like structures and generate field-dependent yield stress. Since the clutch torque is directly governed by the excitation current, the real-time control problem can be formulated as current command generation for haptic rendering.

To execute this strategy, we developed a dedicated bidirectional current driver, as shown in Fig.~\ref{fig:system_architecture}(b). The driving system is centered around an STM32F103 microcontroller (MCU), which functions as a sensor-fusion and control hub. The power stage employs the VNH7040AY driver chip configured in a full H-bridge topology, enabling dynamic reversal of the voltage polarity across the clutch coil for both excitation and demagnetization.

For haptic rendering, the remote interaction force is first converted into a discrete rendering level. Let $F_e(k)$ denote the remote contact force at sampling instant $k$. The nominal rendering level $L(k)$ is defined as
\begin{equation}
L(k)=
\begin{cases}
0, & F_e(k)<F_1,\\
1, & F_1 \le F_e(k)<F_2,\\
2, & F_2 \le F_e(k)<F_3,\\
3, & F_e(k)\ge F_3,
\end{cases}
\label{eq:render_level}
\end{equation}
where $F_1$, $F_2$, and $F_3$ are force thresholds selected for multi-level kinesthetic rendering.

Each rendering level is then mapped to a nominal clutch current,
\begin{equation}
I_{\mathrm{nom}}(k)\in\{I_0,I_1,I_2,I_3\},
\label{eq:nominal_current}
\end{equation}
where $I_0=0$, and $I_1$, $I_2$, and $I_3$ are selected within the validated operating range of the clutch according to the control-oriented current--torque fitting in (\ref{eq:Hill1}). In the current implementation, the clutch is operated in a discrete multi-level rendering mode rather than a fully continuous analog force-rendering mode.

\begin{table}[tb]
    \caption{Key Parameters of the Multi-Level Rendering and Intention-Based Demagnetization Strategy}
    \label{tab:control_parameters}
    \centering
    \small
    \setlength{\tabcolsep}{8pt}
    \renewcommand{\arraystretch}{1.2}
    \begin{tabular}{c c p{0.58\linewidth} p{0.18\linewidth}}
        \toprule
        Symbol & Value & Description\\
        \midrule
        $F_1$ & 1 N & Noise rejection threshold \\
        $F_2$ & 5 N& Threshold between Level 1 and Level 2 \\
        $F_3$ & 10 N& Threshold between Level 2 and Level 3 \\
        $I_0$ & 0 A & Level-0 current (free motion) \\
        $I_1$ & 0.15 A & Level-1 current (low damping) \\
        $I_2$ & 0.30 A & Level-2 current (medium damping)\\
        $I_3$ & 0.50 A & Level-3 current (high damping)\\
        $v_{\mathrm{th}}$ & 0 deg/s & Motion-state threshold\\
        $\Delta F_{\mathrm{th}}$ & 1.98 V & FSR differential threshold\\
        $\alpha$ & 0.5 & Reverse-pulse ratio\\
        $T_d$ & 50 ms & Demagnetization pulse duration\\
        $T_{\mathrm{hold}}$ & 20 ms & consecutive trigger window\\
        Filter & 10 Hz & low-pass cutoff\\
        \bottomrule
    \end{tabular}
\end{table}

\subsection{Intention-Based Demagnetization and Local Override Logic}

Figure~\ref{fig:system_architecture}(c) illustrates the typical haptic interaction scenario, where the master exoskeleton states (A, B, C) correspond synchronously to the slave manipulator states (a, b, c). The control logic consists of a nominal collision-rendering mode and a local override release mode.

The collision-rendering mode is activated when a remote collision is detected and the operator is moving toward the obstacle, i.e.,
\begin{equation}
\mathcal{C}_{\mathrm{col}}(k)=
\bigl(F_e(k)>F_{\mathrm{1}}\bigr)\land\bigl(v(k)>v_{\mathrm{th}}\bigr),
\label{eq:collision_condition}
\end{equation}
where $v(k)$ denotes the IMU-derived motion velocity. Under this condition, the controller applies the nominal current $I_{\mathrm{nom}}(k)$ to render collision resistance.

However, due to the hysteresis and remanence of the MRG clutch, residual damping may persist even after the excitation current is removed. Reverse-pulse demagnetization has been used as an effective engineering approach to suppress residual magnetic effects in magnetically actuated devices~\cite{Senkal2009}. Inspired by this principle, a local intention-based demagnetization strategy is introduced here to mitigate the sticky release effect during disengagement. 
The release mode is triggered when a retraction intention is detected from the handle-side FSRs and the IMU motion state, i.e.,
\begin{equation}
\mathcal{C}_{\mathrm{rel}}(k)=
\bigl(\Delta F_{\mathrm{FSR}}(k)>\Delta F_{\mathrm{th}}\bigr)\land\bigl(v(k)\le v_{\mathrm{th}}\bigr),
\label{eq:release_condition}
\end{equation}
where
\begin{equation}
\Delta F_{\mathrm{FSR}}(k)=FSR_2(k)-FSR_1(k),
\label{eq:fsr_difference}
\end{equation}
and $\Delta F_{\mathrm{th}}$ is the FSR-difference threshold.

Accordingly, the clutch current command is defined as
\begin{equation}
I_{\mathrm{cmd}}(k)=
\begin{cases}
-\alpha I_{\mathrm{nom}}(k), & \mathcal{C}_{\mathrm{rel}}(k)=1,\ 0<t\le T_d,\\
I_{\mathrm{nom}}(k), & \mathcal{C}_{\mathrm{col}}(k)=1,\\
0, & \text{otherwise},
\end{cases}
\label{eq:current_command}
\end{equation}
where $\alpha\in(0,1)$ is the demagnetization reverse-pulse ratio and $T_d$ is the pulse duration. 

The principal control parameters used in the implementation are summarized in Table~\ref{tab:control_parameters}. In implementation, the FSR-difference signal and the IMU-derived velocity were low-pass filtered before threshold comparison. To suppress spurious triggering caused by measurement noise, the release condition was accepted only when the trigger criteria were satisfied for a short consecutive time window.

In other words, the global teleoperation loop generates the nominal current command for multi-level haptic rendering, whereas the local reflex loop can temporarily override this command and inject a short reverse pulse once a release intention is recognized. This design allows the system to preserve stable collision rendering while suppressing the delayed and sticky residual torque during disengagement.

\section{Experiments and Results}

\subsection{Device-Level and System-Level Validation of the Control Strategy}

To validate the control strategy defined in Section~III, both device-level and system-level experiments were conducted. The device-level experiment evaluates whether active demagnetization accelerates clutch release compared with passive power-off and supports the practical selection of the demagnetization parameters. The system-level experiment verifies whether the complete control chain, including multi-level rendering, local intention detection, current override, and torque release, operates as designed.

\subsubsection{Device-Level Validation of Active Demagnetization}

\begin{figure}[tbp]
    \centering
    \includegraphics[width=\linewidth]{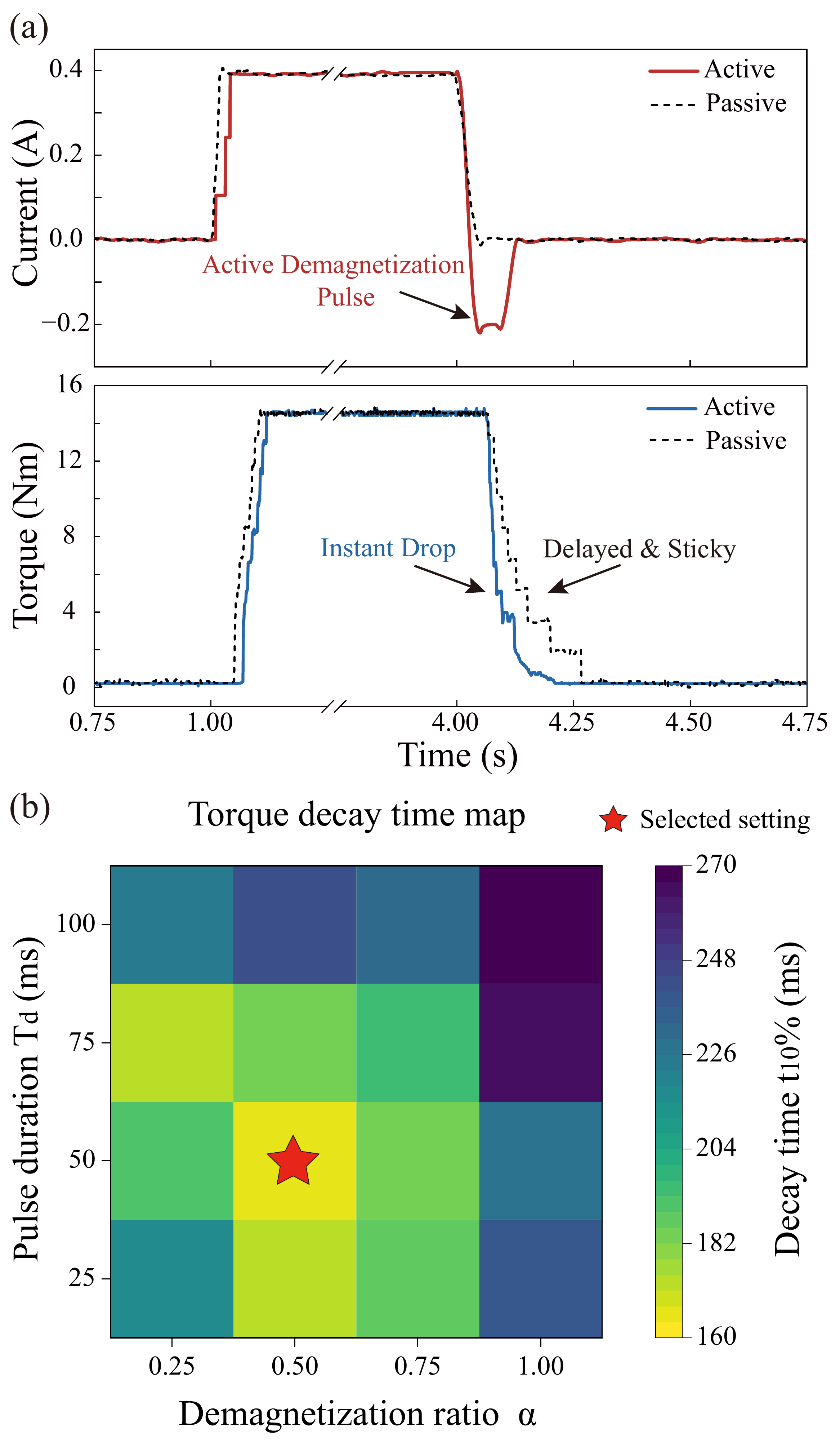}
    \caption{Parameter-sensitivity analysis of the active demagnetization strategy at a current of 0.4 A. (a) transient response comparing passive power-off and active demagnetization at the selected operating point. (b) Torque decay time under different reverse-pulse ratios and pulse durations, with the end point defined as the torque decaying to 10\% of its pre-release value. The marker indicates the parameter pair adopted in the final implementation.}
    \label{fig:demag_scan}
\end{figure}

A device-level bench test was first conducted using the setup shown in Fig.~1(e). Under a nominal excitation current of 0.4\,A, the passive power-off case was obtained by directly removing the excitation current, whereas the active-demagnetization case applied a reverse pulse with a reverse-pulse ratio of 50\% and a duration of 50\,ms. A representative transient comparison at the selected operating point is shown in Fig.~\ref{fig:demag_scan}(a).

Compared with passive power-off, the active demagnetization strategy significantly accelerates torque decay. The reverse pulse suppresses the residual magnetic effect and reduces the torque decay time by approximately 50\% at the selected operating point, thereby effectively reducing the delayed and sticky release behavior associated with passive dissipation.

To further justify the practical parameter choice used in the final implementation, an additional parameter-sensitivity analysis was conducted over the reverse-pulse ratio and pulse duration, as shown in Fig.~\ref{fig:demag_scan}(b). Since the dominant objective of the demagnetization strategy is to shorten the sticky-release interval, the parameter study focuses primarily on the torque decay time as the key selection metric. Based on this analysis, the final demagnetization setting was selected as a compromise between fast release and implementation simplicity.
Among the tested parameter combinations, the setting $(\alpha, T_d)=(0.5, 50\,\mathrm{ms})$ was selected for the final implementation. Although larger reverse-pulse ratios or longer pulse durations can further shorten the torque decay time under some conditions, the selected setting provides a practical compromise among release speed, implementation simplicity, and current-command robustness across the validated operating range.

\subsubsection{System-Level Replay Validation}

Because directly integrating a dynamic torque sensor into wearable joints would degrade transparency and increase structural burden, a two-stage replay-based validation was adopted for system-level evaluation. First, the operator remotely manipulated the exoskeleton while the FSR signals, IMU-derived motion state, and current command were recorded. Once a release intention was detected, the controller injected an active demagnetization pulse according to (\ref{eq:current_command}). Second, the recorded current curves were replayed on the bench platform shown in Fig.~1(e) to measure the corresponding physical torque response of the clutch.

\begin{figure}[htbp]
    \centering
    \includegraphics[width=\linewidth]{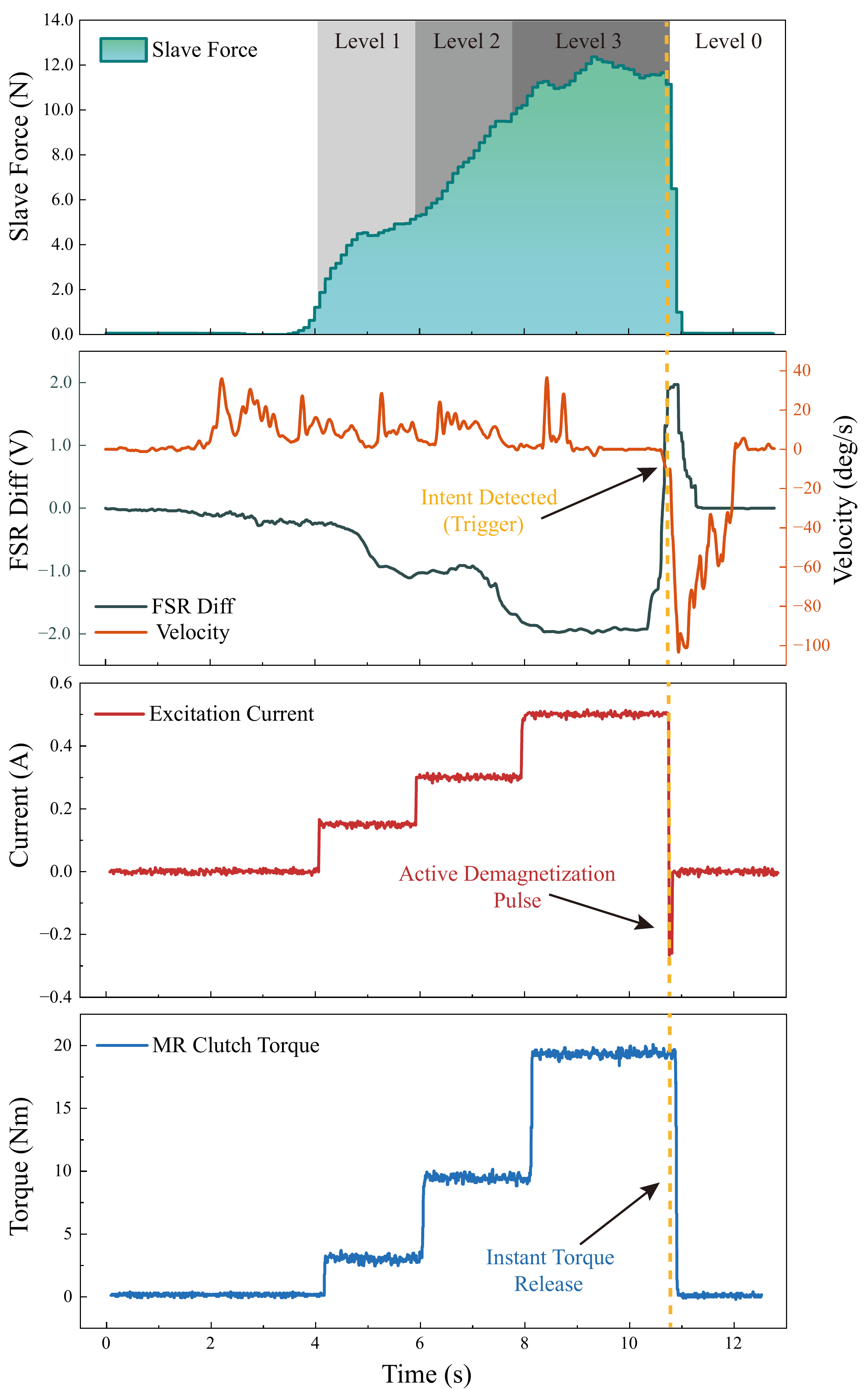}
    \caption{System-level replay validation across a complete collision--holding--release cycle, including slave-side force level, local intention signals, clutch current command, and measured clutch torque.}
    \label{fig:demag_validation}
\end{figure}

The complete synchronized results are shown in Fig.~\ref{fig:demag_validation}. The top panel presents the slave-side force signal, which is quantized into four rendering levels. The second panel shows the local intention-related signals, where a pronounced change in FSR differential together with a non-positive motion state marks the release trigger. The third panel shows the actual current command, including both the nominal multi-level rendering current and the reverse demagnetization pulse. The bottom panel shows the measured clutch torque, which follows the intended multi-level rendering behavior and then drops rapidly to zero once the release pulse is applied.

These results confirm that the control strategy defined in Section~III performs as intended with respect to the command execution and clutch torque response: the nominal loop provides discrete kinesthetic rendering during contact, while the local override loop enables rapid and transparent release once a retraction intention is detected. This replay-based procedure validates the electrical-to-mechanical response of the clutch to the recorded command sequence, but does not fully reproduce the coupled human–exoskeleton impedance during actual wear.

\subsection{Human Factors Study}
Beyond device-level characterization, user-centered perceptual evaluation is important for wearable haptic interfaces, since subjective measures of perceived realism, smoothness, and interaction quality can reveal benefits that are not fully captured by mechanical performance alone~\cite{Maisto2017,Catkin2023}.

To evaluate the perceptual effect of the proposed haptic-rendering and release strategy, a human-subject study was conducted in a virtual teleoperation scenario built on the Unity platform, as shown in Fig. 5(a). The scenario included a robotic arm controlled by the exoskeleton and a small cube with unknown stiffness. As shown in Fig. 5(b), the operator wore the upper-limb exoskeleton and an HTC VIVE PRO 2 head-mounted display. 

At the beginning of each trial, the operator remotely controlled the robotic arm to contact the virtual object, judged its stiffness level based on the rendered kinesthetic sensation, and then retracted from the object. A total of 10 volunteers participated in the study. Two control modes were evaluated, namely passive dissipation and active demagnetization. Each participant completed two trials for each control mode, and each trial involved discrimination among four stiffness levels (Level 0–Level 3). Accordingly, each control mode yielded a total of 80 classification samples (10 participants × 2 trials × 4 levels), which corresponds to the sample counts summarized in the confusion matrices in Fig. 5(c) and Fig. 5(d). To reduce potential order effects, the presentation order of the two control modes was counterbalanced across participants, and a short rest period was provided between successive trials.

\begin{figure*}[tbp]
    \centering
    \includegraphics[width=\linewidth]{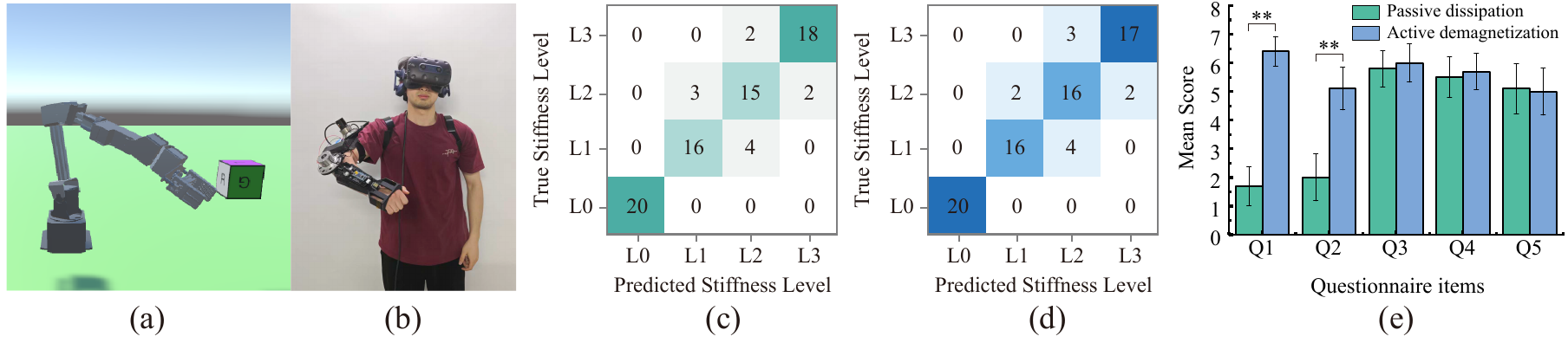}
    \caption{Experimental setup and results of the human factors study. (a) Virtual teleoperation scene in Unity. (b) Participant wearing the exoskeleton and head-mounted display. (c) Confusion matrix under passive dissipation. (d) Confusion matrix under active demagnetization. (e) Questionnaire results reported by the participants after completing the trials under each control mode. Bars represent mean values, and error bars indicate $\pm 1$ SD. Asterisks denote statistical significance based on the Wilcoxon signed-rank test ($ns$: $p>0.05$, *: $p<0.05$, **: $p<0.01$, ***: $p<0.001$). Sample size: $n=10$.}
    \label{fig:human_study}
\end{figure*}

After completing the trials under each control mode, the participants further rated their subjective experience using the questionnaire listed in Table~\ref{tab:questionnaire}. The questionnaire consisted of five items and was scored on a 7-point Likert scale, where 1 indicates “completely disagree” and 7 indicates “completely agree.” For the subjective evaluation, the paired questionnaire scores obtained under the two control modes were compared using the Wilcoxon signed-rank test. Items without significance markings were considered statistically non-significant (ns); specifically, no significant differences were observed for Q3, Q4, or Q5 between passive dissipation and active demagnetization. The mean scores for Q1–Q5 are shown in Fig.~\ref{fig:human_study}(e).

Figure~\ref{fig:human_study}(c) and Fig.~\ref{fig:human_study}(d) show the confusion matrices for passive dissipation and active demagnetization, respectively. For the two extreme cases (Level 0 and Level 3), higher classification accuracy was observed. For the intermediate stiffness levels (Level 1 and Level 2), the passive-dissipation condition achieved classification accuracies of 80\% and 75\%, respectively, while the active-demagnetization condition achieved 80\% for both levels. These results indicate that the system can provide effective kinesthetic cues for objects with sufficiently separated stiffness levels, but also reveal the difficulty of distinguishing subtle intermediate differences.

Importantly, active demagnetization did not increase the overall stiffness-recognition accuracy in the present experiment; both conditions achieved an overall accuracy of 86.25\%. This observation is reasonable because stiffness recognition mainly depends on the collision-rendering phase before disengagement, whereas the sticky effect primarily influences the release phase rather than the initial collision perception.

\begin{table}[htb]
    \caption{Questionnaire Items Used in the Human Factors Study}
    \label{tab:questionnaire}
    \centering
    \small
    \setlength{\tabcolsep}{4pt}
    \begin{tabular}{c p{0.82\linewidth}}
        \toprule
        Q1 & When I tried to move away from the virtual object, the resistance disappeared promptly. \\
        Q2 & The transition from contact to release felt natural and smooth. \\
        Q3 & The haptic feedback clearly helped me perceive the stiffness level of the object. \\
        Q4 & I could decide more easily when to stop pushing and when to retract. \\
        Q5 & I felt noticeable physical effort or arm fatigue during the interaction. \\
        \bottomrule
    \end{tabular}
\end{table}

The questionnaire results show that the active demagnetization condition outperforms passive dissipation in Q1, indicating that the proposed intention-based release strategy improves release promptness. The active-demagnetization condition also achieves a significantly higher score in Q2, demonstrating a smoother and more natural transition from contact to release.

Both control modes receive relatively high scores in Q3 and Q4, suggesting that the participants actively relied on the rendered kinesthetic cues and considered them useful for interaction, particularly when the stiffness levels were sufficiently separated. Combined with the confusion-matrix results, this indicates that active demagnetization does not materially improve the overall stiffness-classification accuracy, but it does reduce cognitive interference during the disengagement stage and improves perceived transparency of release. Therefore, the main perceptual benefit of active demagnetization in the present study lies in improved release transparency rather than enhanced stiffness discrimination accuracy.

The relatively high Q5 scores indicate that the present wearable setup still imposes noticeable physical effort, highlighting the need for further reduction of system weight and improved ergonomics.

\subsection{Teleoperation Task Demonstration}

\begin{figure}[tbp]
    \centering
    \includegraphics[width=\linewidth]{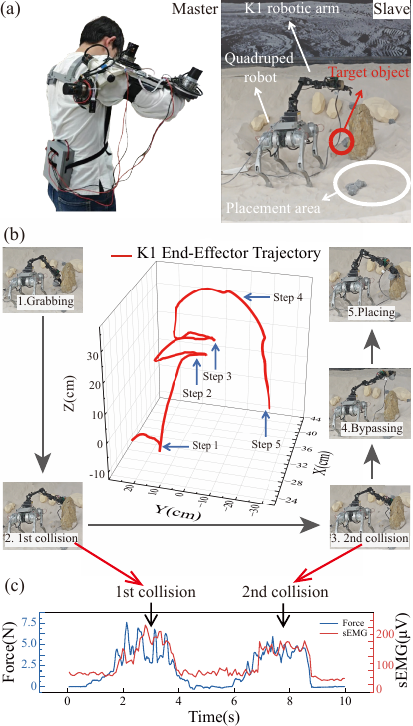}
    \caption{Teleoperation task demonstration in a simulated lunar surface sampling scenario. (a) Master--slave setup, where the operator wears the upper-limb exoskeleton and the slave K1 robotic arm performs remote interaction. (b) K1 end-effector trajectory during grasping, collision, bypassing, and placing. (c) Force and sEMG signals recorded during the two collision events in Supplementary Video~S1.}
    \label{fig:teleop_demo}
\end{figure}

Task-oriented validation is commonly used in teleoperation haptics to assess whether the feedback strategy improves interaction fluency and operator response in representative scenarios~\cite{Girbés2021}. To further demonstrate the feasibility of the proposed haptic-feedback system in a representative teleoperation task, a simulated lunar surface sampling scenario was constructed, as shown in Fig.~\ref{fig:teleop_demo}(a). In this experiment, the operator wore the upper-limb exoskeleton, while the slave side consisted of a K1 robotic arm mounted on a quadruped robot. A force sensor was installed at the K1 end-effector to measure remote contact force. The original K1 gripper was replaced by a dexterous hand (uHand, Hiwonder), which was teleoperated via its companion motion-sensing glove. Further details of the dexterous hand and motion-sensing glove are provided in Supplementary Fig.~S3.

Since the interaction in this task is dominated by elbow motion, an sEMG sensor was placed on the operator's biceps brachii to record muscle activation signals (see Supplementary Fig.~S4). Here, sEMG is used as supportive physiological evidence of the operator's reaction to haptic events, rather than as a primary quantitative measure of haptic fidelity.

During the task, the quadruped robot first transported the K1 arm to the sampling area. The operator then remotely controlled the robotic arm to grasp the target object and finally planned a placement path using the visual information provided by the simulated environment. In realistic lunar sampling scenarios, however, visual information may be incomplete due to environmental occlusion. Under such conditions, collisions may occur during grasping and release. In this case, haptic feedback from the exoskeleton can provide an additional cue for detecting collision events and prompting trajectory replanning.

Supplementary Video~S1 presents a representative demonstration of the task. During the release-path planning phase, the operator encountered two collisions with obstacles, then replanned the trajectory and successfully completed the task by placing the object in the target area. The corresponding end-effector trajectory is shown in Fig.~\ref{fig:teleop_demo}(b).

The force and sEMG signals recorded during the two collision events are shown in Fig.~\ref{fig:teleop_demo}(c). The first collision produced an RMS force of approximately 5.3\,N, with a corresponding RMS sEMG value of about 138\,$\mu$V. The second collision produced an RMS force of approximately 4.2\,N, with a corresponding RMS sEMG value of about 129\,$\mu$V. Although prolonged wearing of the exoskeleton may introduce fatigue and limit the use of sEMG as an absolute quantitative measure, the sharp and localized sEMG peaks occurring around the collision events still indicate that the operator actively reacted to the transmitted haptic cues. Here, sEMG is included only as supplementary physiological evidence indicating the operator’s localized reaction around collision-related haptic events.

In the representative trial summarized in Table 3, active demagnetization reduced the task completion time from 88\,s to 76\,s and the average contact time from 3.71\,s to 2.76\,s, indicating faster recovery from contact and smoother continuation of the task in this demonstration. By contrast, the number of collisions remained unchanged, and the peak contact force showed no clear reduction. These observations suggest that, in the present demonstrated case, the proposed strategy mainly improves release transparency and task continuity rather than collision frequency or collision peak intensity.

Overall, these results demonstrate the task-level feasibility of the proposed system: the operator was able to perceive collision-related haptic feedback, respond to environmental contact, and successfully complete the teleoperation task through visual-haptic coordination.

\begin{table}[htb]
    \caption{Representative task metrics from the demonstrated trial}
    \label{tab:teleop_metrics}
    \centering
    \small
    \setlength{\tabcolsep}{5pt}
    \begin{tabular}{l c c}
        \toprule
        Metric & Passive & Active \\
        \midrule
        Task completion time (s)        & 88   & 76    \\
        Number of collisions            & 2    & 2     \\
        Peak contact force (N)          & 7.87 & 7.52  \\
        Average contact time (s)        & 3.71 & 2.76  \\
        \bottomrule
    \end{tabular}
\end{table}

\section{Discussion}

\subsection{Performance Comparison}

Table~\ref{tab:performance_comparison} compares the proposed clutch with several representative MR clutch and brake designs reported in the literature, using torque-to-mass ratio (TMR), torque-to-volume ratio (TVR), and torque-to-power ratio (TPR) as device-level performance indicators. Under the reported metrics and available published data, the proposed dual-bearing MRG clutch exhibits competitive torque-density characteristics, achieving a TMR of 96.5~N$\cdot$m/kg, a TVR of $3.98\times10^{5}$~N$\cdot$m/m$^{3}$, and a TPR of 4.28~N$\cdot$m/W. These results suggest that the proposed design is capable of delivering substantial locking torque within a compact and wearable form factor.

\begin{table}[htbp]
    \caption{Performance Comparison Among Existing Designs}
    \label{tab:performance_comparison}
    \centering
    \footnotesize
    \setlength{\tabcolsep}{5pt}
    \begin{tabular}{p{2.7cm}<{\centering}p{0.8cm}<{\centering}p{0.8cm}<{\centering}p{0.8cm}<{\centering}p{0.8cm}<{\centering}p{1.0cm}<{\centering}}
        \toprule
        & \parbox{\linewidth}{\centering Our work}
        & \parbox{\linewidth}{\centering MR clutch\cite{Pisetskiy2021}}
        & \parbox{\linewidth}{\centering MR brake\cite{Blake2009}}
        & \parbox{\linewidth}{\centering MR clutch\cite{Onozuka2018}}
        & \parbox{\linewidth}{\centering MR brake\cite{Qin2019}} \\
        \midrule
        Structure                             & Dual-Bearing & Disk       & Drum      & Disk       & Multi-Drum \\
        Max Torque (N$\cdot$m)                & 43.42        & 393        & 0.189     & 5.1        & 1.08 \\
        Volume ($\times10^{-5}$\,m$^{3}$)     & 10.89        & --         & 0.64      & 6.62       & 2.67 \\
        TVR ($\times10^{3}$\,N$\cdot$m/m$^{3}$) & 398.71     & --         & 29.5      & 77.04      & 40.45 \\
        Mass ($\times10^{-1}$\,kg)            & 4.5          & 69         & 0.084     & 2.74       & 2.32 \\
        TMR (N$\cdot$m/kg)                    & 96.5         & 56.9       & 22.5      & 18.61      & 4.66 \\
        Dissipated Power (W)                  & 10.14        & --         & --        & --         & 8.00 \\
        TPR (N$\cdot$m/W)                     & 4.28         & --         & --        & --         & 0.135 \\
        \bottomrule
    \end{tabular}
\end{table}

At the same time, direct comparison across different studies should be interpreted with caution, since the reported performance is affected by differences in device geometry, material composition, magnetic-circuit design, operating conditions, and evaluation protocols. Therefore, Table~\ref{tab:performance_comparison} is intended primarily to provide a coarse device-level performance context rather than a strict one-to-one benchmark under identical conditions.

From a system perspective, the results in Sections~IV-A--IV-C indicate that the practical value of the proposed design lies not only in its torque density, but also in its integration into a wearable haptic-feedback architecture. In particular, the clutch provides sufficient resistive output for multi-level kinesthetic rendering, while the intention-based demagnetization strategy improves release behavior during disengagement. The task-level results further suggest that the main benefit of the proposed strategy is smoother and faster release after contact, rather than a direct reduction in collision occurrence or peak collision intensity in the present demonstrated trial.

\subsection{Limitations and Future Work}

Several limitations of the current system should be acknowledged.

First, part of the validation of the release strategy relied on replay-based bench testing. This approach enables repeatable measurement of the clutch torque response to recorded current commands, and is useful for isolating the electrical-to-mechanical behavior of the device. However, it does not fully reproduce the coupled dynamics of the human arm, the wearable structure, and the local joint impedance during actual operation. As a result, the transient torque perceived by the user during disengagement may differ from that observed in the replay-based measurements. Future work should incorporate in-situ joint-level torque sensing or higher-fidelity estimation methods to better characterize the coupled human--device interaction during wearable use.

Second, the present study validates the proposed system in a discrete multi-level rendering mode rather than in a fully continuous analog force-rendering mode. This design choice is sufficient for demonstrating effective kinesthetic cue delivery and release enhancement, but it also limits the granularity and expressiveness of the rendered haptic feedback. Future work will investigate continuous current-command generation and finer force-to-feedback mapping strategies to improve rendering fidelity.

Third, the intention-based release strategy relies on local FSR measurements together with IMU-derived motion cues. Although this sensing arrangement enables a simple and effective local override mechanism, the FSRs exhibit an activation threshold and a finite dead zone. Consequently, small retraction intentions may not be detected immediately, and a certain amount of residual off-state resistance may still be perceived before the release condition is triggered. Future work will consider richer intention-sensing modalities and more robust sensor-fusion strategies to improve release sensitivity and reduce trigger uncertainty.

Finally, the proposed clutch is based on MRG, whose rheological properties are temperature dependent. While thermal effects were negligible in the relatively short experiments reported here, prolonged high-slip operation may introduce heat accumulation and torque drift. In addition, as a semi-active device, the clutch functions as a programmable resistive element rather than an active torque source, and therefore cannot provide assistive torque to compensate for the weight of the exoskeleton or the user’s arm. These factors may contribute to physical burden during extended operation. Future work will therefore consider thermal compensation, lightweight structural optimization, and hybrid actuation strategies that combine semi-active haptic rendering with active assistance when needed~\cite{Dwivedi2024}.

\section{Conclusion}

This paper presented a wearable haptic-feedback system based on a dual-bearing MRG clutch, together with its interpretive modeling, control, and validation. The proposed clutch was structurally designed, magnetically analyzed, and experimentally characterized, achieving a maximum locking torque of 43.42\,N$\cdot$m at 1.3\,A and a torque-to-mass ratio of 96.5\,N$\cdot$m/kg. These results demonstrate that the proposed device can provide substantial resistive output while maintaining a compact and wearable form factor.

To explain the dominant torque-generation behavior of the device, a physics-inspired interpretive model was introduced to capture the relationship among excitation current, magnetic-field evolution in the bearing gaps, and locking torque generation. Rather than serving as a fully predictive constitutive model, this interpretive model provides physical insight into the monotonic torque growth, saturation behavior, and torque amplification effect introduced by the symmetric dual-bearing structure. For practical implementation, a control-oriented current--torque mapping was used to generate nominal current commands for haptic rendering.

To address the sticky sensation caused by passive bidirectional braking, an intention-based demagnetization strategy was developed using local FSR and IMU cues. Device-level and system-level experiments showed that the reverse-pulse demagnetization method can substantially accelerate torque release compared with passive power-off. Human-subject experiments further showed that active demagnetization does not materially improve the overall stiffness-classification accuracy, but it does improve release promptness and contact-to-release smoothness, thereby reducing perceptual delay during disengagement. Finally, the teleoperation experiments suggest that the proposed strategy improves task fluency mainly by facilitating faster and smoother disengagement after contact, rather than by directly reducing the number of collisions.

Overall, the present results support the proposed dual-bearing MRG clutch as a compact semi-active haptic solution with improved release behavior during disengagement. Future work will focus on continuous rendering, in-situ torque characterization under wearable operation, and more systematic task-level evaluation.

\section*{Conflict of Interest}
The authors declare no conflict of interest.

\nocite{*}
\bibliographystyle{IEEEtran}
\bibliography{references.bib}

\vfill

\end{document}